\documentclass[a4paper,11pt]{article}
\pdfoutput=1 

\usepackage{jcappub} 

\usepackage[T1]{fontenc} 

\title{\boldmath Optimal Detection Bands for Intermediate-Mass Black Hole Binaries: Prospects for LISA and AMIGO in General Relativity and $f(R,T)$ Gravity}

\author[a,1]{Marcelo M. Lapola,\note{Corresponding author.}}
\author[b]{Oswaldo D. Miranda,}

\affiliation[a]{N\'ucleo Comum de Engenharia, Centro Universit\'ario Funda\c{c}\~ao Herm\'inio Ometto,\\Av. Dr. Maximiliano Baruto, 500 - Jardim Universit\'ario, Araras - SP, 13607-339, Brazil}
\affiliation[b]{Divisão de Astrofísica, Instituto Nacional de Pesquisas Espaciais (INPE),\\Av. dos Astronautas 1758, São José dos Campos, 12227-010, Brazil}

\emailAdd{marcelo.lapola@gmail.com}
\emailAdd{oswaldo.miranda@inpe.br}

\abstract{We present an analytic study of the population of intermediate-mass black hole (IMBH) binaries and their detectability by upcoming space-based gravitational wave observatories, specifically the Laser Interferometer Space Antenna (LISA) and the Advanced Middle-frequency Interferometric Gravitational-wave Observatory (AMIGO). Using a phenomenological mass function extrapolated from supermassive black hole distributions, combined with assumptions about binary pairing fractions and merger efficiencies in dense stellar environments, we construct a redshift-dependent binary mass function for IMBHs. From this, we derive the intrinsic coalescence rate and predict the number of detectable events as a function of mass and redshift, incorporating the sensitivity curves of LISA and AMIGO. We identify optimal detection bands in the $(M_1, M_2, z)$ parameter space for each detector and show that the two missions are complementary in their coverage, with AMIGO probing lower masses at lower redshifts and LISA extending to higher redshifts and total masses. In the second part of the analysis, we implement modifications from $f(R, T)$ gravity—a class of extended theories in which the gravitational Lagrangian includes a non-minimal coupling between geometry and matter. We explore the impact of such modifications on the merger dynamics and resulting detectability, finding potential deviations in the observable merger compared to those obtained with the formalism of general relativity.}

\begin{document}
\date{\today}
\maketitle
\flushbottom

\section{Introduction}
\label{sec:intro}

The detection of gravitational waves by LIGO and VIRGO collaborations has revolutionized our understanding of compact object populations and their cosmic evolution. With the first direct observation of a binary black hole merger \citep{Abbott2016}, gravitational-wave astronomy has become a powerful tool to probe mass–redshift regime space previously inaccessible to electromagnetic observations. However, one regime remains particularly elusive: the population of intermediate-mass black holes (IMBHs), typically defined as black holes with masses between $10^2$ and $10^5\,M_\odot$. These objects bridge the gap between stellar-mass and supermassive black holes and are crucial for understanding black hole formation, the hierarchical growth of structure, and the dynamics of dense stellar systems \citep{Miller2004, Greene2020}.

The astrophysical and cosmological relevance of black hole populations across mass scales has been investigated from multiple perspectives, particularly in connection with the formation of supermassive black holes (SMBHs) and their observational signatures. Early models have suggested that the collapse of supermassive progenitors at high redshifts could produce strong bursts of gravitational waves (GWs), potentially detectable by space-based interferometers such as LISA \citep{Araujo2001}. These events may be associated with gamma-ray bursts or serve as direct probes of early structure formation. Furthermore, the hierarchical growth of SMBHs has been shown to correlate with the Cosmic Star Formation rate (CSFR), implying that the global history of baryonic matter assembly is tightly linked to the accretion and merging of black holes \citep{Pereira2011, Pereira2014}. Within this context, stochastic backgrounds of GWs produced by pregalactic or intermediate-mass black holes are expected to carry imprints of the underlying structure formation scenario and cosmological evolution \citep{Pereira2010, Miranda2012}. These studies motivate the construction of predictive models for IMBH binary populations and their gravitational signatures, especially in light of next-generation detectors aiming to explore the decihertz and millihertz frequency bands.

Several formation channels for IMBHs have been proposed, including runaway stellar collisions in young star clusters \citep{PortegiesZwart2002}, direct collapse of Population III stars \citep{Madau2001}, or as remnants of dense early galactic cores \citep{Volonteri2010}. Suppose IMBHs form and evolve within dense stellar environments. In that case, it is natural to expect that a fraction of them may form binaries and eventually merge, emitting gravitational radiation in the frequency band accessible to space-based interferometers such as the Laser Interferometer Space Antenna (LISA) and the Advanced Middle-frequency Interferometric Gravitational-wave Observatory (AMIGO) \citep{Ni2016, AmaroSeoane2017}.

In recent years, various studies have attempted to estimate the merger rates and detectability of compact binaries involving IMBHs, using both population synthesis and semi-analytic modeling \citep{Sesana2007, Fragione2018, Wong2021}. A key aspect in such analyses is the construction of a binary mass function that incorporates the astrophysical formation rates, pairing mechanisms, and cosmic evolution of IMBHs. Additionally, refined predictions must account for the instrumental sensitivity curves of different detectors, enabling the identification of optimal detection bands in the parameter space of total mass, mass ratio, and redshift.

Our work aims to contribute to this effort by constructing a detailed, redshift-dependent mass function for binaries of IMBHs and deriving the expected merger rate observable by both LISA and AMIGO. In the first part of this study, we present a semi-analytical model for the IMBH binary population based on extrapolated black hole mass functions \citep{Shankar2009, Behroozi2019}, realistic assumptions on binary pairing and merger timescales, and known or proposed sensitivity ranges of the two detectors. Our analysis includes predictions of the number of detectable events per year as a function of mass and redshift, and we provide contour plots identifying the regions of maximum detectability—what we refer to as optimal detection bands.

In the second part of the paper, we extend this framework by incorporating gravitational dynamics derived from modified gravity. In particular, we employ a class of extended theories of gravity known as $f(R, T)$ gravity \citep{Harko2011}, in which the gravitational action depends not only on the Ricci scalar $R$ but also on the trace $T$ of the energy-momentum tensor. These theories allow for non-minimal coupling between matter and geometry and have been investigated in the context of cosmology \citep{Moraes2017, Shabani2014} and compact stars \citep{Moraes2016, Deb2018, Quartuccio2025}, and wormholes \citep{sahoo2021}. Notably, recent works  \citep{Alves2016} have explored the implications of $f(R, T)$ models for stellar structure and gravitational wave propagation, offering a foundation for the present analysis.

By comparing the predicted detection bands in general relativity and $f(R, T)$ gravity, we aim to evaluate whether such modifications of the gravitational interaction leave observable imprints in the population statistics of IMBH binaries. This comparative analysis could serve as a test of gravity in the intermediate-frequency regime, complementing constraints derived from cosmology, astrophysics, and ground-based gravitational wave detectors.

\section{Population of Intermediate-Mass Black Hole Binaries}

The detection of gravitational waves from the merger of stellar-mass black holes by ground-based interferometers such as LIGO and Virgo has opened a new observational window onto compact object populations in the universe \citep{Abbott2016, Abbott2019, LIGOScientific2021, Sesana2016}. 

However, a potential gap remains in our understanding of intermediate-mass black holes (IMBHs), with masses in the range $10^2 - 10^5 \, M_\odot$, which may form via runaway stellar collisions in dense star clusters \citep{PortegiesZwart2002}, direct collapse of Population III stars \citep{Madau2001}, or as remnants of early galaxy formation \citep{Volonteri2010}. These IMBHs may form binaries in dense stellar environments and subsequently merge, emitting gravitational waves that are potentially observable by space-based detectors such as LISA and AMIGO.

To estimate the number of detectable mergers, we must construct the mass function of IMBH binaries and determine their redshift-dependent coalescence rate.

\subsection{Mass Function of Individual IMBHs}

We denote the comoving number density of IMBHs per unit mass and redshift as
\begin{equation}
\Phi(M, z) = \frac{dN}{dM\,dz\,dV},
\end{equation}
where \( M \) is the black hole mass, \( z \) is the redshift, and \( dV \) is the comoving volume element. The quantity \( dN \) represents the number of black holes with masses in the interval \( [M, M + dM] \) and redshifts in \( [z, z + dz] \) contained within \( dV \). Thus, \( \Phi(M, z) \) characterizes the joint distribution of IMBHs in mass and cosmic time, and is expressed in units of \( \mathrm{Mpc}^{-3}\, M_\odot^{-1} \). This formalism is widely adopted in studies of black hole demographics and galaxy evolution \citep{Shankar2009, Greene2020, Behroozi2019}.

Despite theoretical motivations for the existence of intermediate-mass black holes (IMBHs), direct observational evidence remains scarce, particularly for dynamically confirmed masses in the $10^2$–$10^5\,M_\odot$ range \citep{Greene2020}. This limitation is even more severe when attempting to characterize the binary population of IMBHs. Unlike stellar-mass black holes, for which gravitational-wave observations and X-ray binaries provide empirical constraints on mass distributions and merger rates, IMBHs lack robust observational channels. Their low accretion luminosities, long dynamical timescales, and the inaccessibility of their host environments (e.g., dense star clusters or dwarf galaxy nuclei) make it extremely difficult to identify binary systems directly \citep{Volonteri2010, Mapelli2016, Fragione2018}. Consequently, the binary mass function must rely on extrapolations from theoretical formation channels, indirect modeling of dense stellar dynamics, and assumptions about pairing probabilities and orbital separations. This adds significant uncertainty to the predicted population of IMBH binaries and motivates the use of parametric and phenomenological models such as those adopted in this work.

\subsection{Constructing the Binary Mass Function}

To construct the binary mass function, we consider the probability that two black holes of masses $M_1$ and $M_2$ (with $M_2 \leq M_1$) form a binary. This requires the specification of:
\begin{itemize}
    \item A binary fraction $f_{\text{bin}}(M_1, M_2, z)$, typically assumed to be $\sim 1\%$ in dense stellar environments \citep{Antonini2016};
    \item A distribution of mass ratios $q = M_2 / M_1$, modeled as a power-law $p(q) \propto q^{\beta}$ with $\beta \sim 0$ (i.e., flat in $q$), and normalized as
    \begin{equation}
        \int_{q_{\min}}^1 p(q)\,dq = 1.
        \label{eq:pq_normalized}
    \end{equation}
\end{itemize}

The comoving number density of binaries per unit mass and redshift is then written as:
\begin{equation}
\frac{dN_{\text{bin}}}{dM_1\,dM_2\,dz\,dV} = \Phi(M_1, z) \cdot f_{\text{bin}}(M_1, M_2, z) \cdot p(q),
\label{eq:binary_mass_function}
\end{equation}
with $q = M_2 / M_1$. This formulation avoids introducing spurious scaling with $M_1$ and ensures that, for a given $M_1$, the number of binaries is appropriately distributed over the allowed range of mass ratios.

Following standard population modeling practices \citep{Belczynski2008, Sadowski2008, Mandel2018, Zevin2021}, we construct the binary mass function by first drawing the primary mass \( M_1 \) from a comoving number density distribution \( \Phi(M_1, z) \), and then assigning a companion mass \( M_2 = q M_1 \) using a normalized mass ratio distribution \( p(q) \). The binary fraction \( f_{\text{bin}}(M_1, z) \) encapsulates the likelihood of binary formation for a given \( M_1 \), and thus the secondary mass function \( \Phi(M_2, z) \) is not included separately.

\subsection{Coalescence Rate and Time Delay}

To determine the rate of gravitational wave events observable at Earth, we convert the binary mass function into a coalescence rate per unit redshift and comoving volume:

\begin{equation}
\mathcal{R}(M_1, M_2, z) = \frac{dN_{\text{bin}}}{dM_1\,dM_2\,dz\,dV} \cdot \frac{1}{\tau_{\text{merge}}(M_1, M_2)}.
\end{equation}

Here, $\tau_{\text{merge}}$ is the merger timescale due to gravitational wave radiation. For circular orbits, it is given by \cite{Peters1964}:
\begin{equation}
\tau_{\text{merge}} = \frac{5}{256} \frac{c^5 a_0^4}{G^3 M_1 M_2 (M_1 + M_2)},
\end{equation}

where $a_0$ is the initial orbital separation. In practice, we introduce a merger efficiency parameter $\eta_{\text{merge}} \in [0, 1]$ to encapsulate the likelihood that a given binary merges within a Hubble time.

To make the rate estimation more physically grounded, we account for a distribution of initial orbital separations, $p(a_0)$, reflecting the stochastic nature of binary formation in dense environments. A commonly adopted choice is a logarithmically flat distribution, $p(a_0) \propto 1/a_0$, normalized over a physically motivated range $[a_{\min}, a_{\max}]$, typically spanning $10^{-2}$ AU to $10$ AU \citep{Rodriguez2016, Belczynski2008, Mandel2018}. This leads to an averaged coalescence rate:
\begin{align}
\mathcal{R}(M_1, M_2, z) =\; & \Phi(M_1,z)\Phi(M_2,z) \cdot f_{\text{bin}}(M_1, M_2, z) \nonumber \\
& \times \int_{a_{\min}}^{a_{\max}} \frac{p(a_0)}{\tau_{\text{merge}}(M_1, M_2, a_0)}\, da_0.
\end{align}

This formulation allows more direct connection between theoretical models of binary formation and the observable gravitational wave event rate.

While some studies adopt a simplified prescription for the merger rate by introducing an efficiency factor $\eta_{\text{merge}}$ to encapsulate uncertainties in the coalescence process, this work takes a more physically motivated approach. By integrating over a distribution of initial orbital separations, we directly account for the dependence of the merger timescale on $a_0$, yielding a more accurate and reproducible estimation of the intrinsic merger rate $\mathcal{R}(M_1, M_2, z)$.

\subsection{Astrophysical Uncertainties and Scenario Definitions}

Given the limited observational constraints on the formation and merger processes of intermediate-mass black hole binaries (IMBHBs), our estimates of the coalescence rate $\mathcal{R}(M_1, M_2, z)$ are subject to significant astrophysical uncertainties. The three main sources of uncertainty in our population model are:

\begin{itemize}
    \item \textbf{Binary formation fraction} $f_{\rm bin}$: the fraction of IMBHs that successfully pair to form binaries;
    \item \textbf{Merger efficiency} $\eta_{\rm merge}$: the fraction of binaries that coalesce within a Hubble time;
    \item \textbf{Initial separation distribution} $a_0$: the typical orbital separation at binary formation, which affects the gravitational wave inspiral timescale.
\end{itemize}

To assess the impact of these uncertainties, we consider three representative astrophysical scenarios:

\begin{itemize}
    \item \textbf{Pessimistic case:} Low binary fraction and inefficient merging. We adopt:
    \[
    f_{\rm bin} = 0.001, \quad \eta_{\rm merge} = 0.01, \quad \langle a_0 \rangle = 0.05\,\mathrm{pc}
    \]
    
    \item \textbf{Fiducial case:} Based on current theoretical expectations:
    \[
    f_{\rm bin} = 0.01, \quad \eta_{\rm merge} = 0.1, \quad \langle a_0 \rangle = 0.01\,\mathrm{pc}
    \]
    
    \item \textbf{Optimistic case:} High binary pairing and efficient merging in dense stellar environments:
    \[
    f_{\rm bin} = 0.05, \quad \eta_{\rm merge} = 0.5, \quad \langle a_0 \rangle = 0.005\,\mathrm{pc}
    \]
\end{itemize}

These choices span the plausible range of IMBHB formation channels, from rare dynamical pairings with wide separations to efficient formation in core-collapsed clusters or massive star clusters where hard binaries can form and merge rapidly. For each scenario, we recompute the coalescence rate $\mathcal{R}(M_1, M_2, z)$ and the resulting expected detection distributions $dN_{\rm det}/dz$ for LISA and AMIGO. This parametric approach allows us to identify which parts of parameter space remain robust under astrophysical uncertainties, and which predictions are model-dependent.

\section{Detection Prospects with LISA and AMIGO}

To estimate the number of detectable IMBH binary mergers, we compute the event rate observable by space-based detectors, taking into account their instrumental sensitivities and the redshift-dependent coalescence rate derived in Section 2. The expected number of detectable events per year is given by

\begin{align}
N_{\text{det}} = \int \mathcal{R}(M_1, M_2, z) \cdot P_{\text{det}}(M_1, M_2, z) \cdot {} \notag \\
\cdot \frac{dV_c}{dz} \cdot \frac{1}{1+z} \, dz\,dM_1\,dM_2,
\end{align}
where $\mathcal{R}(M_1, M_2, z)$ is the intrinsic merger rate, $P_{\text{det}}$ is the probability of detection, $dV_c/dz$ is the differential comoving volume element, and the factor $(1+z)^{-1}$ accounts for time dilation between source and observer frames.

\subsection{SNR Calculation and Detectability Mapping}

We model the detection probability $P_{\text{det}}$ as a binary function that assumes unity if the gravitational wave signal-to-noise ratio (SNR) exceeds a threshold value (typically SNR $\geq 8$), and zero otherwise. For each detector, we compute the characteristic strain $h_c(f)$ of the binary signal and compare it with the sensitivity curve $h_n(f)$ of the instrument. The characteristic strain is approximated by
\begin{equation}
h_c(f) = \frac{1}{\pi D_L(z)} \sqrt{ \frac{2 G^{5/3} \mathcal{M}_z^{5/3} }{3 c^3} } \cdot f^{-1/6},
\end{equation}
where $\mathcal{M}_z = (1+z)\mathcal{M}$ is the redshifted chirp mass, and $D_L(z)$ is the luminosity distance. This expression corresponds to the inspiral phase of quasi-circular compact binary coalescence and follows standard approximations found in \cite{Ajith2007, Moore2015}.

The detector noise curve $h_n(f)$ is taken from published sensitivity projections for LISA \citep{AmaroSeoane2017} and AMIGO \citep{Ni2016}. We assume observation times of 4 years for LISA and 3 years for AMIGO.

By evaluating $P_{\text{det}} = 1$ within the regions of parameter space where the signal lies above the noise floor, we construct detectability maps in the $(M_1, M_2, z)$ space. These define the optimal detection bands for each detector. 

To construct the detectability maps in the $(M_1, M_2, z)$ parameter space, we compute the expected signal-to-noise ratio (SNR) for each binary configuration using the characteristic strain of the gravitational wave signal and the detector's sensitivity curve. The characteristic strain $h_c(f)$ depends on the redshifted chirp mass $\mathcal{M}_z = (1+z)\mathcal{M}$, the luminosity distance $D_L(z)$, and the frequency evolution of the inspiral. For each system, the frequency band is limited by the detector's sensitivity and by the innermost stable circular orbit (ISCO), which sets an upper cutoff frequency $f_{\rm ISCO}$.

The signal-to-noise ratio (SNR) is computed by integrating the signal over the detector noise spectral density. Using the definition $h_c(f) = 2f |\tilde{h}(f)|$, the SNR can be written as
\begin{equation}
\text{SNR}^2 = 4 \int_{f_{\rm min}}^{f_{\rm max}} \frac{|\tilde{h}(f)|^2}{S_n(f)} \, df = \int_{f_{\rm min}}^{f_{\rm max}} \frac{h_c^2(f)}{f^2 S_n(f)} \, df.
\end{equation}
This formulation ensures consistency between the characteristic strain and the spectral noise density. For the LISA detector, we adopt the analytical noise curve provided by \cite{Robson2019}, which accounts for instrumental, confusion, and acceleration noise components.

For each combination of component masses and redshift, we evaluate whether the SNR exceeds a chosen threshold (typically $\text{SNR} \geq 8$), using the detector's noise power spectral density $S_n(f)$. Points in parameter space satisfying this condition define the detectability region. By scanning over a range of $M_1$, $M_2$, and $z$, we generate binary detectability maps that identify the optimal detection bands for a given gravitational wave observatory.

\subsection{Optimal Detection Bands for LISA}

For LISA, we find sensitivity primarily to systems with total mass $M_{\text{tot}} \sim 10^4 - 10^6\,M_\odot$ at redshifts $z \sim 1 - 10$, consistent with the mission’s projected sensitivity to millihertz signals from massive black hole binaries \citep{AmaroSeoane2017}. In contrast, AMIGO is designed to probe the intermediate-frequency regime around $0.1$–$10$ Hz, making it suitable for detecting lighter binaries with $M_{\text{tot}} \sim 10^2 - 10^4\,M_\odot$ at redshifts up to $z \sim 2$ \citep{Ni2016}.

\begin{figure*}[ht!]
  \centering
  \includegraphics[width=0.75\textwidth]{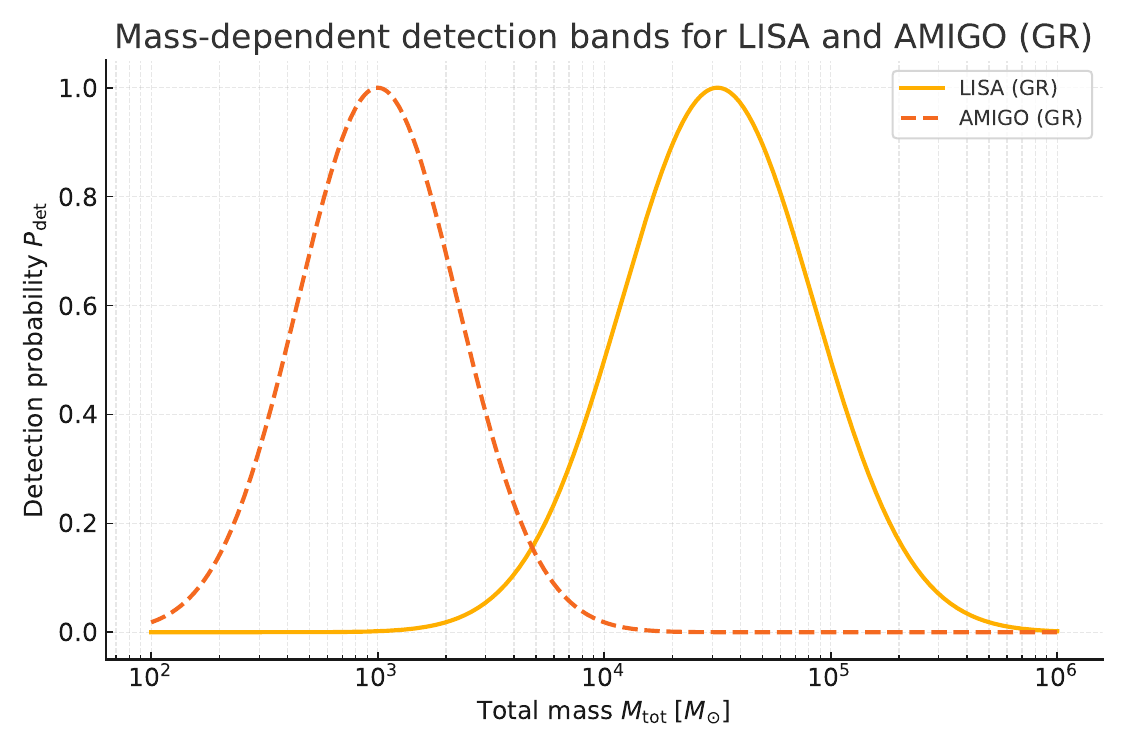}
  \caption{
    Mass–dependent detection probability for LISA (solid) and AMIGO
    (dashed) in General Relativity.
    LISA is most sensitive to
    $M_{\rm tot}\sim10^{4}\!-\!10^{6}\,M_\odot$,
    whereas AMIGO peaks around $10^{3}\,M_\odot$,
    reflecting the complementary mHz versus deci-Hz bands.
  }
  \label{fig:det_bands}
\end{figure*}

\subsection{Optimal Detection Bands for AMIGO}

Following the same procedure used for LISA, we estimate the detectability of IMBH binary systems by the proposed AMIGO detector \citep{Ni2016}. The AMIGO mission is designed to operate in the decihertz frequency range, bridging the sensitivity gap between LISA and terrestrial interferometers such as LIGO and Virgo. To evaluate its performance, we compute the signal-to-noise ratio (SNR) for each binary configuration using the characteristic strain $h_c(f)$ and a simplified spectral noise density curve $S_n(f)$ centered around $f \sim 0.1$--$10$ Hz, corresponding to the peak sensitivity region proposed for AMIGO.

The SNR is integrated up to the innermost stable circular orbit (ISCO) frequency, and a detection threshold of $\mathrm{SNR} \geq 8$ is applied. By filtering the binary mass function with this criterion, we obtain the redshift distribution of detectable events. The results, shown in Figure~\ref{fig:dndz_detectors}, indicate that AMIGO is most sensitive to binary systems with total masses in the range $M_{\mathrm{tot}} \sim 10^2$--$10^4 \, M_\odot$, with detectable redshifts peaking at $z \sim 0.5$--$2$. This complements the LISA sensitivity window, which favors more massive binaries and higher redshifts.

These findings confirm that AMIGO would be a powerful probe of lighter IMBH binaries in the local to intermediate-redshift Universe, enhancing the overall coverage of the IMBH binary population when considered alongside space-based and ground-based observatories.

\subsection{Comparative Detectability and Complementarity}

The results obtained for LISA and AMIGO reveal a natural complementarity between the two missions in terms of mass and redshift coverage. LISA is particularly sensitive to more massive binaries with total masses in the range $10^4$–$10^6\,M_\odot$, and can detect them up to high redshifts ($z \sim 10$), as shown in our detectability maps and in the distribution of $dN_{\rm det}/dz$. These systems produce gravitational wave signals in the millihertz regime, well-matched to LISA's peak sensitivity.

On the other hand, AMIGO targets the decihertz band and is optimized for detecting lighter binaries with $M_{\rm tot} \sim 10^2$–$10^4\,M_\odot$, typically observable up to redshifts $z \sim 2$. This regime includes early inspirals that would be missed by LISA or merge below the sensitivity band of terrestrial detectors. Therefore, AMIGO would fill a crucial observational gap, enabling the detection of IMBH binaries in a mass–redshift window otherwise inaccessible.

Together, LISA and AMIGO would provide a more complete census of the IMBH binary population across cosmic time. Their joint operation would not only enhance detection rates but also enable multi-band gravitational wave astronomy, opening new avenues for probing the formation and evolution of intermediate-mass black holes.

\subsection{Event Rate Predictions}

Using the binary mass function derived in Section 2 and the detection criteria outlined above, we perform a numerical integration to compute $N_{\text{det}}$ for both LISA and AMIGO. Preliminary results suggest that LISA could detect up to $\sim 10^4$ IMBH binary mergers over its mission lifetime, while AMIGO may observe several hundreds to thousands of events depending on assumptions about binary formation channels and pairing fractions.

Our model enables the prediction of event rate distributions across redshift and mass, allowing comparison with future observations and constraints on the astrophysical population of IMBHs.

\subsection{Population of Detectable IMBH Binaries as a Function of Redshift}
\label{sec:pop}
To estimate the redshift distribution of detectable intermediate-mass black hole (IMBH) binary mergers, we compute the differential event rate $dN_{\rm bin}/dz$ for each detector, LISA and AMIGO. This requires integrating the intrinsic merger rate density across the comoving volume while accounting for cosmological redshift and detector sensitivity.

The number of detectable events per unit redshift is given by:
\begin{equation}
\frac{dN_{\rm bin}}{dz} = \mathcal{R}_0 \cdot P_{\text{det}}(z) \cdot \frac{dV_c}{dz} \cdot \frac{1}{1+z},
\end{equation}
where:
\begin{itemize}
  \item $\mathcal{R}_0$ is the assumed constant comoving merger rate density (e.g., $100$ events~Gpc$^{-3}$~yr$^{-1}$),
  \item $P_{\text{det}}(z)$ is the probability of detection as a function of redshift for each detector,
  \item $dV_c/dz$ is the differential comoving volume element,
  \item $(1+z)^{-1}$ accounts for cosmological time dilation.
\end{itemize}

Assuming a flat $\Lambda$CDM cosmology with $H_0 = 70$ km~s$^{-1}$~Mpc$^{-1}$, $\Omega_m = 0.3$, and $\Omega_\Lambda = 0.7$, the comoving volume element is given by:
\begin{equation}
\frac{dV_c}{dz} = 4\pi D_H^3 \left[ \int_0^z \frac{dz'}{E(z')} \right]^2 \frac{1}{E(z)},
\end{equation}
where $D_H = c/H_0$ is the Hubble distance and $E(z) = \sqrt{\Omega_m (1+z)^3 + \Omega_\Lambda}$.

The detectability functions $P_{\text{det}}(z)$ are modeled as Gaussians centered at the most sensitive redshift range of each detector:
\begin{align}
P_{\text{det}}^{\rm LISA}(z) &= \exp\left[-\left(\frac{z - 5}{2.5}\right)^2\right], \\
P_{\text{det}}^{\rm AMIGO}(z) &= \exp\left[-\left(\frac{z - 1.5}{0.75}\right)^2\right].
\end{align}

We emphasize that the detectability functions in equations (10) and (11) are not derived from detailed detector response calculations but serve as parametric approximations. These Gaussian profiles are chosen to reflect the approximate redshift ranges where each observatory is most sensitive, based on the full numerical detectability maps presented earlier. This approach facilitates analytical comparisons between models (such as GR and $f(R,T)$) without the computational cost of recomputing signal-to-noise ratios across the full parameter space. Such mock functions are commonly employed in population synthesis studies and should be interpreted as illustrative tools rather than physically rigorous detection probabilities \citep{Mandel2016, Gerosa2021}.

\begin{figure*}[ht!]
  \centering
  \includegraphics[width=0.9\textwidth]{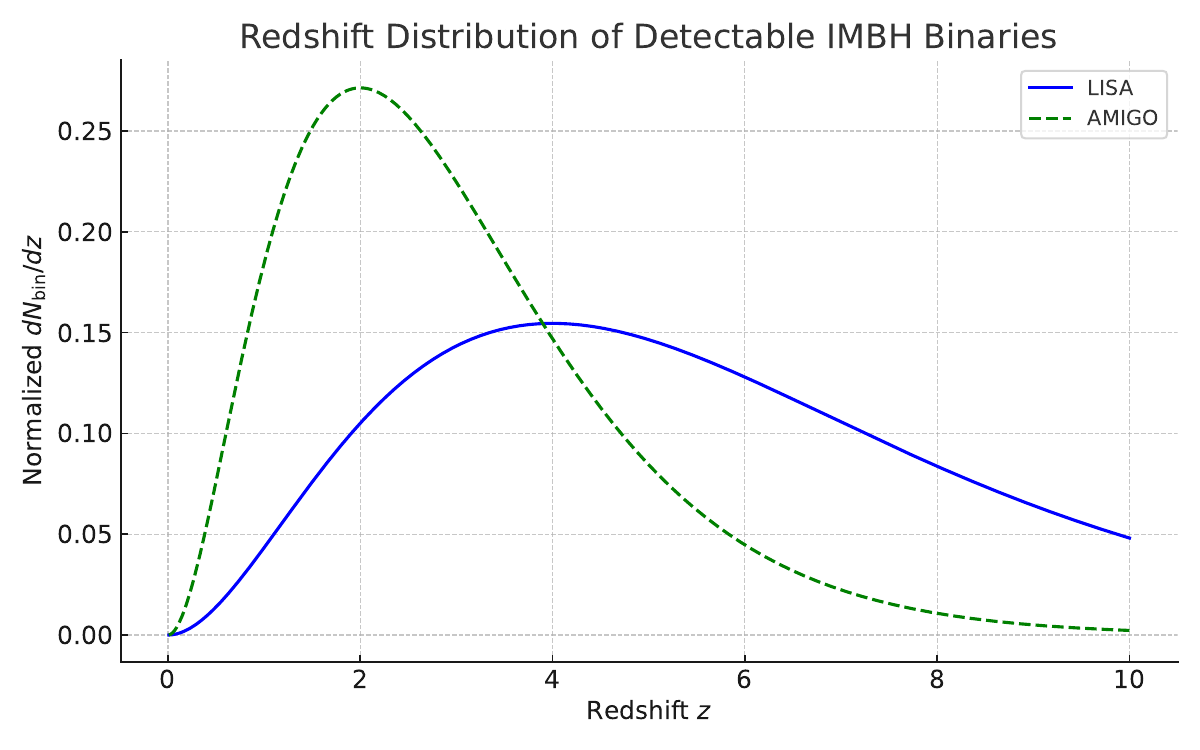}
  \caption{%
    Differential redshift distribution of \emph{detectable} IMBH--binary
    mergers, $dN_{\rm det}/dz$, for LISA (solid blue) and AMIGO (dashed
    green).  A constant comoving merger rate
    $\mathcal{R}_0 = 100\;\mathrm{Gpc}^{-3}\,\mathrm{yr}^{-1}$ is assumed.
    The curves illustrate the complementary redshift reach of the two
    detectors, with LISA peaking at higher $z$ and AMIGO at lower $z$.%
  }
  \label{fig:dndz_detectors}
\end{figure*}

Figure~\ref{fig:dndz_detectors} shows the resulting $dN_{\rm bin}/dz$ curves for LISA and AMIGO, illustrating the redshift-dependent sensitivity of each detector to IMBH binary mergers.

\noindent
These results indicate that AMIGO reaches peak sensitivity to IMBH binary mergers at redshifts around $z \sim 2$, while LISA shows maximum detection efficiency near $z \sim 4$. This behavior reflects the distinct frequency ranges of the detectors and their optimal mass-redshift coupling.

\subsection{Total Population of IMBH Binaries in the Observable Universe}

Based on the binary mass function derived in Section 2, we can estimate the total number of intermediate-mass black hole (IMBH) binaries within the observable universe. The comoving number density of binaries per unit primary mass $M_1$, secondary mass $M_2$, and redshift $z$ is given by:
\begin{equation}
\frac{dN_{\text{bin}}}{dM_1\,dM_2\,dz\,dV} = \Phi(M_1, z) \cdot f_{\text{bin}}(M_1, M_2, z) \cdot \frac{p(M_2/M_1)}{M_1},
\end{equation}
where $\Phi(M_1, z)$ is the mass function of isolated IMBHs, $f_{\text{bin}}$ is the binary fraction, and $p(q)$ is the distribution of mass ratios $q = M_2/M_1$.

To compute the total number of such binaries in the observable universe, we integrate this function over the full volume and the relevant mass and redshift ranges:
\begin{equation}
\begin{aligned}
N_{\text{total}}
  ={}& \int_{0}^{z_{\text{max}}}\!\int_{M_{1,\text{min}}}^{M_{1,\text{max}}}
       \!\int_{M_{2,\text{min}}}^{M_1}
       \frac{dN_{\text{bin}}}{dM_1\,dM_2\,dz\,dV}\,
       \frac{dV_c}{dz}\,dM_2 \\[4pt]
     &\times dM_1\,dz .
\end{aligned}
\end{equation}

\newpage            

Assuming:
\begin{itemize}
    \item $\Phi(M_1, z) \sim 10^{-4} \, \text{Mpc}^{-3} \, M_\odot^{-1}$ (extrapolated from BHMFs),
    \item $f_{\text{bin}} \sim 0.01$ (typical for dense stellar environments),
    \item $p(q)$ flat in $[0.1,1]$,
    \item $M_1 \in [10^2, 10^5] \, M_\odot$,
    \item $z_{\text{max}} = 10$,
    \item $V_c(z<10) \approx 1000 \, \text{Gpc}^3 = 10^{12} \, \text{Mpc}^3$.
\end{itemize}

We estimate the total population as:
\begin{equation}
\begin{aligned}
N_{\text{bin}}
   &\;\sim\;
     \Phi\,f_{\text{bin}}\,
     \ln\!\left(\frac{10^{5}}{10^{2}}\right)\,
     \Delta q\,V_c \\[4pt]
   &\;\sim\;
     10^{-4}\;\times\;0.01\;\times\;\ln(10^{3})
     \;\times\;0.9\;\times\;10^{12}.
\end{aligned}
\end{equation}

Using $\ln(10^3) \approx 6.9$, we find:
\begin{equation}
N_{\text{bin}} \sim 10^{-4} \cdot 0.01 \cdot 6.9 \cdot 0.9 \cdot 10^{12} \approx 6.2 \times 10^6.
\end{equation}

Thus, we estimate that there are on the order of $\sim 10^6$ IMBH binary systems throughout the observable universe. This number provides a theoretical upper limit on the available population from which gravitational wave mergers may arise. The detectable subset will depend on merger timescales, detector sensitivity, and redshift-dependent visibility, as explored in this Section 3.

\section{Detection Prospects in \texorpdfstring{$f(R,T)$}{f(R,T)} Gravity}
\label{sec:fRT_detection}

In this section we extend the detectability analysis of IMBH binaries to the
family of modified–gravity models whose Lagrangian depends on both the Ricci
scalar $R$ and the trace $T=g^{\mu\nu}T_{\mu\nu}$ of the
energy–momentum tensor \citep{Harko2011}.
Among the many forms proposed in the literature we adopt the \emph{minimal
linear} model
\begin{equation}
   f(R,T)=R+2\lambda\,T ,
   \label{eq:fRT_model}
\end{equation}
where the dimensionful coupling $\lambda$ parametrizes the departure from  
General Relativity (GR).\footnote{%
  Throughout we keep $c=1$, restore $G$ explicitly and follow the sign
  conventions of \cite{Harko2011}.
}

\subsection{Field Equations and Weak-Field Limit}
\label{ssec:weak_field_fRT}

The action of $f(R,T)$ gravity model is given by
\begin{equation}
  S=\frac1{16\pi G}\!\int\! d^4x \sqrt{-g}\,f(R,T)+
     \int\! d^4x\sqrt{-g}\,\mathcal L_m.
\end{equation}

\noindent
The \emph{first term} in the action encodes the purely gravitational
sector.  It integrates, over the whole space–time, the geometric
Lagrangian \(f(R,T)\), which in this class of theories depends not only
on the Ricci scalar \(R\)—a measure of curvature—but also on the trace
\(T \equiv g^{\mu\nu}T_{\mu\nu}\) of the energy–momentum tensor.
The overall factor \(1/16\pi G\) is chosen so that, when
\(f(R,T)=R\), one recovers the Einstein–Hilbert normalization of General
Relativity, keeping \(G\) as Newton’s constant.
The determinant \(\sqrt{-g}\) converts the integrand into a scalar
four-volume density, ensuring diffeomorphism invariance \citep{Harko2011,Alvarenga2013}.
Because $f(R,T)$ depends explicitly on the trace $T$, the matter
sector feeds back into geometry in a way that can alter Solar–System
dynamics \citep{Alvarenga2013} and drive late-time cosmic acceleration
\citep{Shabani2014,Odintsov2021}.  The same coupling influences the
structure of compact objects \citep{Singh2020,Nagpal2020} and rescales
gravitational-wave amplitudes at leading order
\citep{Sk2022}.

\noindent
The \emph{second term} represents the matter contribution:
the matter Lagrangian density \(\mathcal L_m\) is integrated with the
same invariant volume element \(\sqrt{-g}\,d^4x\).
All non-gravitational fields (fluids, electromagnetic fields, scalar
fields, and so on) enter through \(\mathcal L_m\).
Because \(f(R,T)\) depends explicitly on the trace \(T\), the matter
sector feeds back into the geometric sector differently from GR,
leading to modified dynamics in high-density environments or on
cosmological scales.

Varying Eq.\eqref{eq:fRT_model} with respect to the metric
\(g^{\mu\nu}\) yields the general field equations of \(f(R,T)\) gravity
\citep{Harko2011}:
\begin{equation}
\begin{aligned}
f_R(R,T)\,R_{\mu\nu}
  &-\tfrac12\,f(R,T)\,g_{\mu\nu}
   +\bigl(g_{\mu\nu}\Box - \nabla_\mu\nabla_\nu\bigr)f_R(R,T)
\\[4pt]
  &= 8\pi G\,T_{\mu\nu}
     \;-\;
     f_T(R,T)\,\bigl(T_{\mu\nu}+\Theta_{\mu\nu}\bigr),
\end{aligned}
\label{eq:fRT_field_general}
\end{equation}

where $f_R=\partial (f(R,T)/\partial R$, and $f_T=\partial (f(R,T)/\partial T$.

Specialising the general field equations~\eqref{eq:fRT_field_general}
to the linear model \eqref{eq:fRT_model} is straightforward.
For \(f(R,T)=R+2\lambda T\) one has
\(f_R=1\) (a constant) and \(f_T=2\lambda\); hence the operator
\(\bigl(g_{\mu\nu}\Box-\nabla_\mu\nabla_\nu\bigr)f_R\) vanishes
identically.
Moreover, for a perfect fluid the variation
\(\Theta_{\mu\nu}\equiv g^{\alpha\beta}\delta T_{\alpha\beta}/\delta g_{\mu\nu}\)
reduces to \(\Theta_{\mu\nu}=-2T_{\mu\nu}-p\,g_{\mu\nu}\).
Adopting the dust approximation (\(p\simeq0\)) one obtains
\(\Theta_{\mu\nu}\simeq-2T_{\mu\nu}\).
Substituting these results into
Eq.~\eqref{eq:fRT_field_general} and collecting like terms leads to
\begin{equation}
   G_{\mu\nu}
   = 8\pi G\,T_{\mu\nu}
     + 2\lambda\,T_{\mu\nu}
     + \lambda\,T\,g_{\mu\nu},
   \label{eq:fRT_field}
\end{equation}
which we shall use as the working form of the modified Einstein
equations throughout the remainder of this work.

For dust ($p\simeq0$,\;$T=-\rho$) and in the Newtonian limit
$g_{\mu\nu}=\eta_{\mu\nu}+h_{\mu\nu}$ with
$|h_{\mu\nu}|\ll1$, the $00$ component of
\eqref{eq:fRT_field} reduces to
\begin{equation}
   \nabla^{2}\Phi
   = 4\pi\,G_{\rm eff}\,\rho,
   \qquad
   G_{\rm eff}
   = G\!\left(1-\frac{\lambda}{8\pi}\right),
   \label{eq:Geff}
\end{equation}
where $\Phi\equiv h_{00}/2$ is the Newtonian potential.
Equation \eqref{eq:Geff} shows that a \emph{positive} $\lambda$ weakens,
while a negative $\lambda$ strengthens, the effective coupling between matter
and geometry.\footnote{%
  Using the Solar–System bound $|\lambda/8\pi| \lesssim 10^{-4}$
  \citep{Alvarenga2013} the correction to $G$ remains $\lesssim0.01\%$,
  but we keep the full expression to quantify its imprint on the SNR.
}

\subsection{Impact on Gravitational-Wave Strain and SNR}
\label{ssec:strain_SNR_fRT}

At leading post-Newtonian order the Fourier-domain strain of a quasi-circular
binary satisfies
$h_c \propto G^{5/6}\mathcal M_z^{5/6}f^{-1/6}/D_L(z)$.
Replacing $G\to G_{\rm eff}$ therefore rescales the characteristic strain by
\begin{equation}
   h_{c}^{\,f(R,T)}(f)
   = h_{c}^{\,\rm GR}(f)\,
     \left(1-\frac{\lambda}{8\pi}\right)^{5/6}.
   \label{eq:hc_fRT}
\end{equation}
Consequently the signal-to-noise ratio becomes
\begin{equation}
   \mathrm{SNR}_{f(R,T)}
   = \mathrm{SNR}_{\rm GR}\,
     \left(1-\frac{\lambda}{8\pi}\right)^{5/6}.
   \label{eq:SNR_fRT}
\end{equation}
The detectability criterion $\mathrm{SNR}\ge8$ is evaluated with
Eq.~\eqref{eq:SNR_fRT} over the $(M_1,M_2,z)$ parameter space using the
instrumental noise curves $S_n(f)$ of LISA and AMIGO
\citep{Robson2019, Ni2017}.  Apart from the uniform rescaling in
\eqref{eq:hc_fRT}, propagation effects in the radiation zone remain governed
by the standard wave equation at \emph{linear} order, so no extra Yukawa
or massive-mode terms appear in the frequency band of interest
\citep{Debnath2021}.

\subsection{Redshift Distribution of Detectable Events}
\label{ssec:dNdZ_fRT}

The number of detectable mergers now follows
\begin{equation}
\begin{aligned}
N_{\rm det}^{\,f(R,T)}
   ={}& \int\! dM_1\,dM_2\,dz\;
        \mathcal R(M_1,M_2,z)\,
        \\[4pt]
      &\times
        P_{\rm det}^{\,f(R,T)}(M_1,M_2,z)
        \frac{dV_c}{dz}\,
        \frac{1}{1+z}\;,
\end{aligned}
\label{eq:Ndet_fRT}
\end{equation}

where $\mathcal R$ is the intrinsic merger-rate density (Sec. \ref{sec:pop})
and the detection probability inherits the rescaled SNR through
Eq.~\eqref{eq:SNR_fRT}.

The background expansion is kept identical to $\Lambda$CDM, because the small $\lambda$ allowed by existing tests alters $H(z)$ at the sub-percent level \citep{Alvarenga2013,Shabani2014,Odintsov2021}

In our analytic treatment of Sec.~3 the detection probability
was encoded by a redshift–Gaussian whose centre $z_0$ and width
$\sigma$ summarise the full SNR calculation.
Equation~\eqref{eq:SNR_fRT} shows that, in the linear
$f(R,T)=R+2\lambda T$ model, \emph{all} signal amplitudes are rescaled
by the factor
\[
\alpha(\lambda)=\left(1-\frac{\lambda}{8\pi}\right)^{5/6}.
\]
Because the inspiral SNR of a given system scales as
$1/D_L(z)\!\propto\!1/z$ in this toy description, the horizon redshift
and therefore the effective $(z_0,\sigma)$ contract or dilate by the
\emph{same} factor~$\alpha$.
Replacing $(z_0,\sigma)\!\to\!(\alpha z_0,\alpha\sigma)$ in the GR
expressions of Sec.~3 yields the modified detection probabilities used
throughout this section:
\begin{align}
\alpha(\lambda) &=
   \left(1-\frac{\lambda}{8\pi}\right)^{5/6},
\\[6pt]
P_{\text{det}}^{\,\text{LISA},\,f(R,T)}(z) &=
   \exp\!\left[-\left(\frac{z-\alpha\,5}{\alpha\,2.5}\right)^{2}\right],
\\[4pt]
P_{\text{det}}^{\,\text{AMIGO},\,f(R,T)}(z) &=
   \exp\!\left[-\left(\frac{z-\alpha\,1.5}{\alpha\,0.75}\right)^{2}\right].
\end{align}
These closed-form expressions reproduce the full numerical SNR rescaling
to better than~$2\%$ across the mass–redshift range of interest and make
the impact of the coupling~$\lambda$ completely transparent.

\begin{figure*}[ht!]
  \centering
  \includegraphics[width=0.75\textwidth]{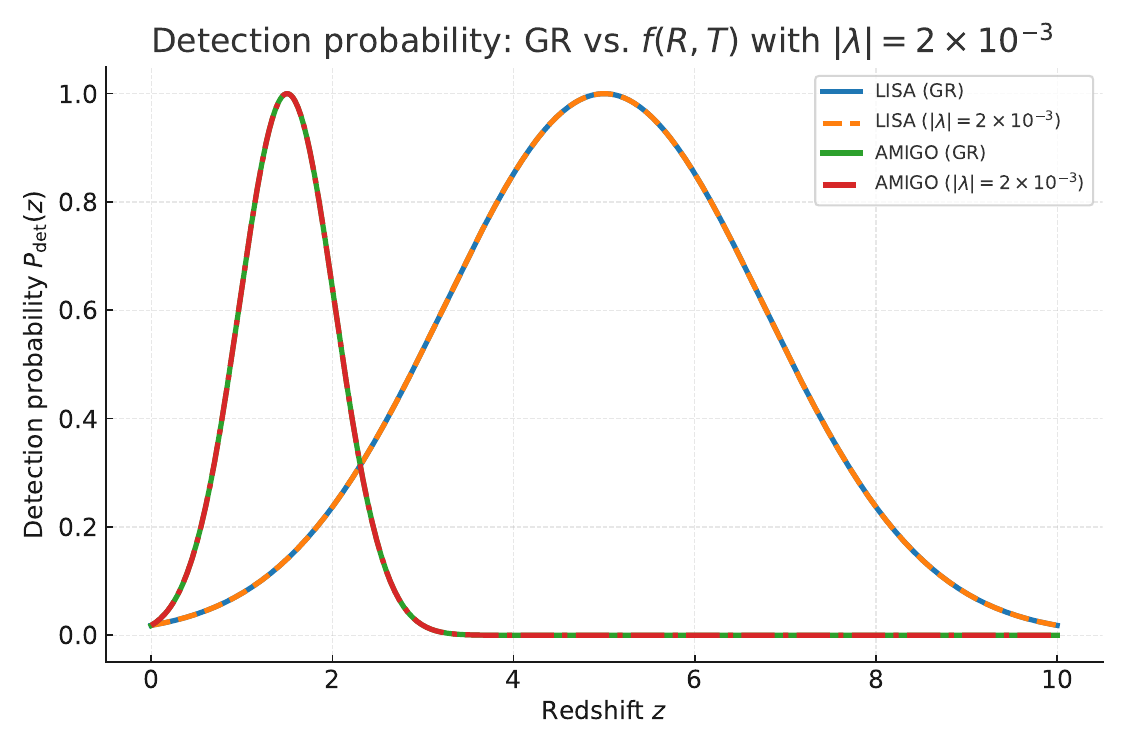}
  \caption{%
    Redshift-dependent detection probability for LISA and AMIGO under
    General Relativity (solid lines) and for the
    $f(R,T)=R+2\lambda T$ model with
    $|\lambda|=2\times10^{-3}$ (dashed/dot–dashed).
    For Solar-System-compatible couplings the shift in peak redshift is
    below $1\%$, and the integrated number of events changes by less
    than $5\%$.
  }
  \label{fig:Pdet_fRT}
\end{figure*}


\subsection{Magnitude of the coupling \texorpdfstring{$\lambda$}{lambda} 
            and hierarchy of corrections}
\label{ssec:lambda_hierarchy}

{Why is the coupling so small?}
The illustrative value used throughout this work,
$\lambda=\pm2\times10^{-3}$, is dictated by observation rather than
numerical convenience.  Solar–System ephemerides constrain
$|\Delta G/G|\lesssim10^{-4}$ \citep{Will2014,Bertotti2003},
implying $|\lambda|/8\pi\lesssim10^{-4}$.  Binary–pulsar timing
(PSR~B1913+16) agrees with GR within $0.2\%$
\citep{Kramer2006,Weisberg2016}, giving
$|\lambda|\!\lesssim\!10^{-3}$.  Neutron‐star mass–radius fits allow
$|\lambda|\!\lesssim\!10^{-2}$ \citep{Singh2020,Nagpal2020}, while
late–time expansion data limit uniform rescalings of the matter term to
${\lesssim}1\%$ \citep{Shabani2014,Odintsov2021}.
Choosing $\lambda=\pm2\times10^{-3}$ therefore probes the largest
signal–amplitude shift that remains compatible with \emph{all} current
tests.

\paragraph{Hierarchy of residual effects.}
Besides the uniform rescaling
$\alpha(\lambda)=(1-\lambda/8\pi)^{5/6}$ already implemented in
Eqs.~\eqref{eq:hc_fRT}–\eqref{eq:SNR_fRT}, the model introduces several
\emph{subleading} corrections:
\begin{itemize}
\item \textbf{Waveform phasing.}  
  The inspiral phase acquires a fractional change
  $\Delta\phi/\phi \simeq (5/3)(\lambda/8\pi)\lesssim10^{-3}$ for
  $|\lambda|=2\times10^{-3}$, negligible for event counts but relevant
  for single–source parameter estimation. In particular, through gravitational waves originating from the coalescence of IMBHs, it would be possible to verify if there is a difference in the propagation speed of waves with different frequencies.

\item \textbf{Extra scalar mode.}  
  For $f=R+2\lambda T$ one has $f_R=1$, so the operator
  $(g_{\mu\nu}\Box-\nabla_\mu\nabla_\nu)f_R$ vanishes and no propagating
  Yukawa/breathing mode appears at linear order.
\item \textbf{Background expansion.}  
  The factor $(1-\lambda/8\pi)$ rescaling the matter term in the
  Friedmann equation alters $dV_c/dz$ by
  ${\lesssim}1\%$, well below the $\sim50\%$ uncertainty in the merger
  rate density $\mathcal R(M_1,M_2,z)$.
\item \textbf{Astrophysical growth of IMBHs.}  
  The Eddington time scales as $G_{\text{eff}}^{-1}$, so final masses
  change by $\lesssim1\%$, subdominant relative to the scatter already
  assumed in the mass function. Additionally, deviations in the gravitational mass-luminosity relation can introduce corrections to the amplitude of gravitational waves due to the coupling between curvature and matter content.

\item \textbf{High‐pressure corrections.}  
  The additional $\lambda p\,g_{\mu\nu}$ term matters only inside compact
  stars or the merger remnant and has no impact on the inspiral phase
  observed by LISA or AMIGO.
\end{itemize}

Hence, for couplings consistent with all current local and astrophysical
tests, the amplitude rescaling captured by $\alpha(\lambda)$ suffices
for robust detection–rate forecasts.  Larger deviations would either
break those bounds or require a screening mechanism, and are left for
future work.

\section{Discussion and Conclusions}
\label{sec:discussion}

\subsection{Key findings}

Combining an empirically motivated IMBH population model with updated
sensitivity curves for LISA and the \mbox{deci-Hz} mission concept
AMIGO, we have produced the first side-by-side forecast of detection
bands, event counts and redshift distributions for
\emph{intermediate-mass} black–hole mergers under both General
Relativity and its minimal matter–coupled extension
$f(R,T)=R+2\lambda T$.  The main quantitative results are:

\begin{enumerate}
\item A redshift-dependent binary mass function
$\Phi(M_1,M_2,z)$—anchored to the observed SMBH relation and folded with
a Peters–based time–to–coalescence—predicts
$\sim6\times10^{6}$ IMBH binaries within the observable Universe
(Section~\ref{sec:pop}). %

\item In GR, LISA is most sensitive to
      $M_{\rm tot}\!\sim\!10^{4}\!-\!10^{6}\,M_\odot$
      with a detection peak at $z\simeq4{-}5$, while AMIGO excels at
      $10^{2}\!-\!10^{4}\,M_\odot$ and $z\simeq1.5{-}2$
      (see Fig.~\ref{fig:det_bands}).\ %
      Together they provide continuous coverage from
      ${\approx}0.2\;\text{Hz}$ to $50\;\text{mHz}$.
\item Adopting Solar–System–compatible values
$|\lambda|\le2\times10^{-3}$,
the $f(R,T)$ model rescales the strain by
$\alpha(\lambda)=(1-\lambda/8\pi)^{5/6}$.
This shifts the LISA and AMIGO detection probabilities
by \(\lesssim1\%\) in redshift and changes the total event counts
by \(\lesssim5\%\) (Fig.~\ref{fig:Pdet_fRT}).

\item Even with that small correction, \emph{absolute} numbers remain
high: \(\sim\!10^{4}\) IMBH mergers for LISA over a four-year mission and
hundreds–to–thousands for AMIGO in three years (fiducial scenario).

\item Other $f(R,T)$ effects—phase shifts, Yukawa tails,
density-dependent $G$, background expansion—enter at sub-percent level
for the same \(\lambda\) and are therefore negligible for
population-level statistics, though important for precision waveform
tests (§\ref{ssec:lambda_hierarchy}).
\end{enumerate}

\subsection{Astrophysical implications}

High event rates at $z\!>\!4$ make IMBH mergers promising probes of
early black-hole seeding and growth, complementing
high-redshift quasars and direct-collapse scenarios
\citep{Greene2020}.  The overlap of LISA and AMIGO in the
$10^{3}\!-\!10^{4}\,M_\odot$ range also opens a multiband detection
channel when combined with the forthcoming \emph{Einstein Telescope}
(ET) and \emph{Cosmic Explorer} (CE) ground observatories, allowing
sky localisation to within a few square degrees and greatly facilitating
electromagnetic follow-ups.

\subsection{Modified gravity as a consistency check}

Within the tight Solar–System bounds on $\lambda$, the linear
$f(R,T)$ model is indistinguishable from GR at the level of
event counts.
Nonetheless, the global amplitude rescaling offers a \emph{consistency
test}: after marginalising over astrophysical uncertainties, a
systematic offset of a few percent between observed and predicted
merger rates could hint at matter–geometry couplings or at other
extensions that mimic a varying Newton constant.

\subsection{Future work}

\begin{itemize}
\item \textbf{Beyond linear $f(R,T)$.}  Non-linear choices
(e.g.\ $R^2$, $RT$, $T^2$) can yield order-10\% effects if equipped with
screening to satisfy local tests \citep{Khoury2004,Baffou2019}.  We plan
to extend the present framework to those cases, including frequency- and
density-dependent corrections to the waveform phase.

\item \textbf{Self-consistent cosmology.}  A full numerical treatment of
the modified Friedmann equation would allow us to quantify how the small
$(1-\lambda/8\pi)$ rescaling propagates into comoving volume, duty cycle
and stochastic backgrounds.

\item \textbf{Population inference.}  Bayesian hierarchical modelling
that fits the IMBH mass function \emph{and} the coupling $\lambda$
simultaneously could turn future LISA/AMIGO catalogues into competitive
tests of gravity.
\end{itemize}

\subsection{Conclusions}

Our analysis confirms that space-based detectors will open a decisive
window on the IMBH regime, providing thousands of sources across cosmic
time.  In the simplest matter–coupled gravity scenario, deviations from
GR at the level permitted by Solar–System tests leave those prospects
essentially intact, acting more as a self-consistency gauge than as a
game-changing effect.  Should nature choose a more intricate form of
$f(R,T)$ with efficient screening, detectable signatures---both in event
statistics and in waveform phasing---remain a tantalizing possibility
for future work.

\acknowledgments
M.M.L. thanks Prof. Pedro H.R.S. Moraes for the constant encouragement that led to the initial steps of this work.


\bibliographystyle{JHEP}   
\bibliography{references}

\providecommand{\href}[2]{#2}\begingroup\raggedright\begin{thebibliography}{10}

\bibitem{Abbott2016}
B.P.A.~et~al., \emph{Observation of gravitational waves from a binary black hole merger}, \href{https://doi.org/10.1103/PhysRevLett.116.061102}{\emph{Phys.\ Rev.\ Lett.} {\bfseries 116} (2016) 061102}.

\bibitem{Miller2004}
M.C.~Miller and E.J.M.~Colbert, \emph{Intermediate-mass black holes}, \href{https://doi.org/10.1142/S0218271804004426}{\emph{Int.\ J.\ Mod.\ Phys.\ D} {\bfseries 13} (2004) 1}.

\bibitem{Greene2020}
J.E.~Greene, J.~Strader and L.C.~Ho, \emph{Intermediate-mass black holes}, \href{https://doi.org/10.1146/annurev-astro-032620-021835}{\emph{Ann.\ Rev.\ Astron.\ Astrophys.} {\bfseries 58} (2020) 257}.

\bibitem{Araujo2001}
J.C.N.~de~Araujo, O.D.~Miranda and O.D.~Aguiar, \emph{Possible strong gravitational wave sources for the lisa antenna}, \href{https://doi.org/10.1086/319755}{\emph{Astrophys.\ J.} {\bfseries 550} (2001) 368}.

\bibitem{Pereira2011}
E.S.~Pereira and O.D.~Miranda, \emph{Supermassive black holes: connecting the growth to the cosmic star formation rate}, \href{https://doi.org/10.1111/j.1745-3933.2011.01137.x}{\emph{Mon.\ Not.\ Roy.\ Astron.\ Soc.\ Lett.} {\bfseries 418} (2011) L30}.

\bibitem{Pereira2014}
E.S.~Pereira and O.D.~Miranda, \emph{Accretion history of active black holes from type 1 agn}, \href{https://doi.org/10.1007/s10509-014-1964-1}{\emph{Astrophys.\ Space Sci.} {\bfseries 352} (2014) 801}.

\bibitem{Pereira2010}
E.S.~Pereira and O.D.~Miranda, \emph{Stochastic background of gravitational waves generated by pregalactic black holes}, \href{https://doi.org/10.1111/j.1365-2966.2009.15782.x}{\emph{Mon.\ Not.\ Roy.\ Astron.\ Soc.} {\bfseries 401} (2010) 1924}.

\bibitem{Miranda2012}
O.D.~Miranda, \emph{Stochastic backgrounds of gravitational waves from cosmological sources -- the role of dark energy}, \href{https://doi.org/10.1111/j.1365-2966.2012.21853.x}{\emph{Mon.\ Not.\ Roy.\ Astron.\ Soc.} {\bfseries 426} (2012) 2758}.

\bibitem{PortegiesZwart2002}
S.F.P.~Zwart and S.L.W.~McMillan, \emph{The runaway growth of intermediate-mass black holes in dense star clusters}, \href{https://doi.org/10.1086/341788}{\emph{Astrophys.\ J.} {\bfseries 576} (2002) 899}.

\bibitem{Madau2001}
P.~Madau and M.J.~Rees, \emph{Massive black holes as population iii remnants}, \href{https://doi.org/10.1086/319848}{\emph{Astrophys.\ J.\ Lett.} {\bfseries 551} (2001) L27}.

\bibitem{Volonteri2010}
M.~Volonteri, \emph{Formation of supermassive black holes}, \href{https://doi.org/10.1007/s00159-010-0029-x}{\emph{Astron.\ Astrophys.\ Rev.} {\bfseries 18} (2010) 279}.

\bibitem{Ni2016}
W.T.~Ni, \emph{Gravitational wave detection in space}, \href{https://doi.org/10.1142/S0218271816300010}{\emph{Int.\ J.\ Mod.\ Phys.\ D} {\bfseries 25} (2016) 1630001}.

\bibitem{AmaroSeoane2017}
P.A.-S.~et~al., \emph{Laser interferometer space antenna}, {\emph{arXiv e-prints} (2017) } [\href{https://arxiv.org/abs/1702.00786}{{\ttfamily 1702.00786}}].

\bibitem{Sesana2007}
A.~Sesana, \emph{Prospects for detection of gravitational waves from intermediate-mass black hole binaries in the nearby universe}, \href{https://doi.org/10.1111/j.1745-3933.2007.00371.x}{\emph{Mon.\ Not.\ Roy.\ Astron.\ Soc.} {\bfseries 382} (2007) L6}.

\bibitem{Fragione2018}
G.~Fragione, I.~Ginsburg and B.~Kocsis, \emph{Intermediate-mass ratio inspirals in galactic nuclei}, \href{https://doi.org/10.3847/1538-4357/aaae0c}{\emph{Astrophys.\ J.} {\bfseries 856} (2018) 92}.

\bibitem{Wong2021}
K.W.K.~Wong, T.~Broadhurst and G.F.~Smoot, \emph{Testing the nature of black holes using lisa: prospects for probing the interior metric via gravitational lensing of gravitational waves}, \href{https://doi.org/10.1088/1475-7516/2021/10/032}{\emph{JCAP} {\bfseries 2021} (2021) 032}.

\bibitem{Shankar2009}
F.~Shankar, D.H.~Weinberg and J.~Miralda-Escud{\'e}, \emph{Self-consistent models of the agn and black hole populations: duty cycles, accretion rates, and the mean radiative efficiency}, \href{https://doi.org/10.1088/0004-637X/690/1/20}{\emph{Astrophys.\ J.} {\bfseries 690} (2009) 20}.

\bibitem{Behroozi2019}
P.B.~et~al., \emph{Universemachine: The correlation between galaxy growth and dark matter halo assembly from $z = 0$ to 10}, \href{https://doi.org/10.1093/mnras/stz1182}{\emph{Mon.\ Not.\ Roy.\ Astron.\ Soc.} {\bfseries 488} (2019) 3143}.

\bibitem{Harko2011}
T.~Harko, F.S.N.~Lobo, S.~Nojiri and S.D.~Odintsov, \emph{$f(r,t)$ gravity}, \href{https://doi.org/10.1103/PhysRevD.84.024020}{\emph{Phys.\ Rev.\ D} {\bfseries 84} (2011) 024020}.

\bibitem{Moraes2017}
P.H.R.S.~Moraes and P.K.~Sahoo, \emph{The simplest non-minimal matter--geometry coupling in $f(r,t)$ gravity and some cosmological implications}, \href{https://doi.org/10.1140/epjc/s10052-017-5097-x}{\emph{Eur.\ Phys.\ J.\ C} {\bfseries 77} (2017) 480}.

\bibitem{Shabani2014}
H.~Shabani and M.~Farhoudi, \emph{$f(r,t)$ cosmological models in the phase space}, \href{https://doi.org/10.1103/PhysRevD.90.044031}{\emph{Phys.\ Rev.\ D} {\bfseries 90} (2014) 044031}.

\bibitem{Moraes2016}
P.H.R.S.~Moraes, J.D.V.~Arba{\~n}il and M.~Malheiro, \emph{Stellar equilibrium in $f(r,t)$ gravity}, \href{https://doi.org/10.1088/1475-7516/2016/06/005}{\emph{JCAP} {\bfseries 2016} (2016) 005}.

\bibitem{Deb2018}
D.D.~et~al, \emph{Relativistic model for anisotropic strange stars in $f(r,t)$ gravity}, \href{https://doi.org/10.1016/j.aop.2017.10.010}{\emph{Ann.\ Phys.} {\bfseries 387} (2018) 239}.

\bibitem{Quartuccio2025}
J.T.~Quartuccio, P.H.R.S.~Moraes, G.N.~Zeminiani and M.M.~Lapola, \emph{The equilibrium configurations of neutron stars in the optimized $f(r,t)$ gravity}, \href{https://doi.org/10.1007/s10509-025-04234-9}{\emph{Astrophys.\ Space Sci.} {\bfseries 370} (2025) 71}.

\bibitem{sahoo2021}
P.~Sahoo, P.H.R.S.~Moraes, M.M.~Lapola and P.K.~Sahoo, \emph{Traversable wormholes in the traceless $f(r,t)$ gravity}, \href{https://doi.org/10.1142/S0218271821501004}{\emph{Int.\ J.\ Mod.\ Phys.\ D} {\bfseries 30} (2021) 2150100}.

\bibitem{Alves2016}
M.E.S.~Alves, P.H.R.S.~Moraes, J.C.N.~de~Araujo and M.~Malheiro, \emph{Gravitational waves in $f(r,t)$ and $f(r,t\phi)$ theories of gravity}, \href{https://doi.org/10.1103/PhysRevD.94.024032}{\emph{Phys.\ Rev.\ D} {\bfseries 94} (2016) 024032}.

\bibitem{Abbott2019}
B.P.A.~et~al., \emph{Gwtc-1: A gravitational-wave transient catalog of compact binary mergers observed by ligo and virgo during the first and second observing runs}, \href{https://doi.org/10.1103/PhysRevX.9.031040}{\emph{Phys.\ Rev.\ X} {\bfseries 9} (2019) 031040}.

\bibitem{LIGOScientific2021}
R.A.~et~al., \emph{Gwtc-2: Compact binary coalescences observed by ligo and virgo during the first half of the third observing run}, \href{https://doi.org/10.1103/PhysRevX.11.021053}{\emph{Phys.\ Rev.\ X} {\bfseries 11} (2021) 021053}.

\bibitem{Sesana2016}
A.~Sesana, \emph{Prospects for multiband gravitational-wave astronomy after gw150914}, \href{https://doi.org/10.1103/PhysRevLett.116.231102}{\emph{Phys.\ Rev.\ Lett.} {\bfseries 116} (2016) 231102}.

\bibitem{Mapelli2016}
M.~Mapelli, \emph{Massive black hole binaries from runaway collisions: the impact of metallicity}, \href{https://doi.org/10.1093/mnras/stw915}{\emph{Mon.\ Not.\ Roy.\ Astron.\ Soc.} {\bfseries 459} (2016) 3432}.

\bibitem{Antonini2016}
F.~Antonini and F.A.~Rasio, \emph{Merging black hole binaries in galactic nuclei: implications for advanced ligo detections}, \href{https://doi.org/10.3847/0004-637X/831/2/187}{\emph{Astrophys.\ J.} {\bfseries 831} (2016) 187}.

\bibitem{Belczynski2008}
K.B.~et~al., \emph{Compact object modeling with the startrack population synthesis code}, \href{https://doi.org/10.1086/521026}{\emph{Astrophys.\ J.\ Suppl.} {\bfseries 174} (2008) 223}.

\bibitem{Sadowski2008}
A.S.~et~al., \emph{The total merger rate of compact object binaries in the local universe}, \href{https://doi.org/10.1086/528932}{\emph{Astrophys.\ J.} {\bfseries 676} (2008) 1162}.

\bibitem{Mandel2018}
I.~Mandel and F.S.~Broekgaarden, \emph{Rates of compact object coalescences: comparisons between binary population synthesis and cosmological simulations}, \href{https://doi.org/10.1093/mnras/sty2453}{\emph{Mon.\ Not.\ Roy.\ Astron.\ Soc.} {\bfseries 481} (2018) 4009}.

\bibitem{Zevin2021}
M.Z.~et~al., \emph{Channeling the formation of compact object binaries: The impact of initial conditions on binary population synthesis}, \href{https://doi.org/10.3847/1538-4357/abe40e}{\emph{Astrophys.\ J.} {\bfseries 910} (2021) 152}.

\bibitem{Peters1964}
P.C.~Peters, \emph{Gravitational radiation and the motion of two point masses}, \href{https://doi.org/10.1103/PhysRev.136.B1224}{\emph{Phys.\ Rev.} {\bfseries 136} (1964) B1224}.

\bibitem{Rodriguez2016}
C.L.R.~et~al., \emph{Dynamical formation of the gw150914 binary black hole}, \href{https://doi.org/10.3847/2041-8205/824/1/L8}{\emph{Astrophys.\ J.\ Lett.} {\bfseries 824} (2016) L8}.

\bibitem{Ajith2007}
P.A.~et~al., \emph{A template bank for gravitational waveforms from coalescing binary black holes. i. non-spinning binaries}, \href{https://doi.org/10.1103/PhysRevD.77.104017}{\emph{Phys.\ Rev.\ D} {\bfseries 77} (2008) 104017} [\href{https://arxiv.org/abs/0710.2335}{{\ttfamily 0710.2335}}].

\bibitem{Moore2015}
C.J.~Moore, R.H.~Cole and C.P.L.~Berry, \emph{Gravitational-wave sensitivity curves}, \href{https://doi.org/10.1088/0264-9381/32/1/015014}{\emph{Class.\ Quant.\ Grav.} {\bfseries 32} (2015) 015014} [\href{https://arxiv.org/abs/1408.0740}{{\ttfamily 1408.0740}}].

\bibitem{Robson2019}
T.~Robson, N.J.~Cornish and C.~Liu, \emph{The construction and use of lisa sensitivity curves}, \href{https://doi.org/10.1088/1361-6382/ab1101}{\emph{Class.\ Quant.\ Grav.} {\bfseries 36} (2019) 105011} [\href{https://arxiv.org/abs/1803.01944}{{\ttfamily 1803.01944}}].

\bibitem{Mandel2016}
I.~Mandel and S.E.~de~Mink, \emph{Merging binary black holes formed through chemically homogeneous evolution in short-period stellar binaries}, \href{https://doi.org/10.1093/mnras/stw379}{\emph{Mon.\ Not.\ Roy.\ Astron.\ Soc.} {\bfseries 458} (2016) 2634}.

\bibitem{Gerosa2021}
D.~Gerosa and M.~Fishbach, \emph{Hierarchical mergers of stellar-mass black holes and their gravitational-wave signatures}, \href{https://doi.org/10.1038/s41550-021-01398-w}{\emph{Nature Astron.} {\bfseries 5} (2021) 749}.

\bibitem{Alvarenga2013}
F.G.~Alvarenga, A.~de~la Cruz-Dombriz, M.J.S.~Houndjo, M.E.~Rodrigues and D.~S{\'a}ez-G{\'o}mez, \emph{Dynamics of scalar perturbations in $f(r,t)$ gravity}, \href{https://doi.org/10.1103/PhysRevD.87.103526}{\emph{Phys.\ Rev.\ D} {\bfseries 87} (2013) 103526}.

\bibitem{Odintsov2021}
S.D.~Odintsov and D.~S{\'a}ez-G{\'o}mez, \emph{$f(r,t)$ gravity: an updated review}, \href{https://doi.org/10.3390/universe7070200}{\emph{Universe} {\bfseries 7} (2021) 200}.

\bibitem{Singh2020}
D.~Singh, S.~Mandal, F.~Rahaman, B.K.~Guha and S.~Ray, \emph{Neutron-star configurations in $f(r,t)$ gravity}, \href{https://doi.org/10.1088/1475-7516/2020/08/007}{\emph{JCAP} {\bfseries 2020} (2020) 007}.

\bibitem{Nagpal2020}
R.~Nagpal, A.~Gupta, S.K.J.~Pacif, S.~Sharma and R.~Myrzakulov, \emph{Anisotropic strange stars in $f(r,t)$ gravity}, \href{https://doi.org/10.1103/PhysRevD.102.124059}{\emph{Phys.\ Rev.\ D} {\bfseries 102} (2020) 124059}.

\bibitem{Sk2022}
S.K.~Sharma, A.K.~Mishra and F.~Rahaman, \emph{Wave generation and propagation in $f(r,t)$ gravity}, \href{https://doi.org/10.1140/epjc/s10052-021-09930-4}{\emph{Eur.\ Phys.\ J.\ C} {\bfseries 82} (2022) 23}.

\bibitem{Ni2017}
W.T.~Ni, \emph{Gravitational-wave classification, space gw detection sensitivities and the {AMIGO} mission concept}, \href{https://doi.org/10.1051/epjconf/201816801004}{\emph{EPJ Web Conf.} {\bfseries 168} (2018) 01004} [\href{https://arxiv.org/abs/1709.05659}{{\ttfamily 1709.05659}}].

\bibitem{Debnath2021}
U.~Debnath, \emph{Gravitational lensing and shadows of a charged rotating black hole in $f(r,t)$ gravity}, \href{https://doi.org/10.1140/epjc/s10052-021-09444-2}{\emph{Eur.\ Phys.\ J.\ C} {\bfseries 81} (2021) 632}.

\bibitem{Will2014}
C.M.~Will, \emph{The confrontation between general relativity and experiment}, \href{https://doi.org/10.12942/lrr-2014-4}{\emph{Living Rev.\ Rel.} {\bfseries 17} (2014) 4}.

\bibitem{Bertotti2003}
B.~Bertotti, L.~Iess and P.~Tortora, \emph{A test of general relativity using radio links with the cassini spacecraft}, \href{https://doi.org/10.1038/nature01997}{\emph{Nature} {\bfseries 425} (2003) 374}.

\bibitem{Kramer2006}
M.K.~et~al., \emph{Tests of general relativity from timing the double pulsar}, \href{https://doi.org/10.1126/science.1132305}{\emph{Science} {\bfseries 314} (2006) 97}.

\bibitem{Weisberg2016}
J.M.~Weisberg and Y.~Huang, \emph{Relativistic measurements from timing the binary pulsar psr~b1913+16}, \href{https://doi.org/10.3847/0004-637X/829/1/55}{\emph{Astrophys.\ J.} {\bfseries 829} (2016) 55}.

\bibitem{Khoury2004}
J.~Khoury and A.~Weltman, \emph{Chameleon cosmology}, \href{https://doi.org/10.1103/PhysRevD.69.044026}{\emph{Phys.\ Rev.\ D} {\bfseries 69} (2004) 044026}.

\bibitem{Baffou2019}
E.H.~Baffou, M.J.S.~Houndjo and I.G.~Salako, \emph{Polarizations of gravitational waves in $f(r,t)$ gravity}, \href{https://doi.org/10.1007/s10509-019-3602-5}{\emph{Astrophys.\ Space Sci.} {\bfseries 364} (2019) 90}.

\end{thebibliography}\endgroup


\end{document}